\documentclass[12pt]{iopart}
\usepackage{graphicx}

\begin{document}

\title[ ]{Systematic studies of binding energy
dependence of neutron - proton momentum correlation function}

\author{Y. B. Wei\dag \ddag, Y. G. Ma\dag, W. Q.
Shen\dag, G. L. Ma\dag, K. Wang\dag, X. Z. Cai\dag, C. Zhong\dag,
W. Guo\dag, J. G. Chen\dag, D. Q. Fang\dag, \\ W. D. Tian\dag, X.
F. Zhou \footnote[1]{To whom Correspondence should be addressed:
ygma@sinap.ac.cn}}

\address{\dag\ Shanghai Institute of Applied Physics, Chinese Academy of Sciences,
  P. O. Box 800-204, Shanghai 201800, China}
\address{\ddag\ Graduate School of the Chinese Academy of Sciences,
China}

\begin{abstract}
Hanbury Brown-Twiss (HBT) results of the neutron-proton
correlation function have been systematically investigated for a
series nuclear reactions with light projectiles with help of
Isospin-Dependent Quantum Molecular Dynamics model. The
relationship between the binding energy per nucleon of the
projectiles and the strength of the neutron-proton HBT  at small
relative momentum has been obtained. Results show that
neutron-proton HBT results are sensitive to the binding energy per
nucleon.
\end{abstract}






\section{INTRODUCTION}
The intensity interferometry was developed by Hanbury-Brown and
Twiss \cite{Brown,Goldhaber,Koonin} in the 1950s, as a means of
determining the dimension of distant astronomical objects. After a
number of terrestrial experiments were performed to confirm this
technique, the method of intensity interferometry is now commonly
referred to as the Hanbury-Bworn/Twiss (HBT) effect. HBT
interferometry differs from ordinary amplitude interferometry in
that it comprises intensities, rather than amplitudes, at
different points. It indicates the effects of Bose or Fermi
statistics even if the phase of the (light or matter) wave is
disturbed by uncontrollable random fluctuations or if the counting
rate is very low \cite{Boal}. It has also become an important tool
in high energy heavy ion collision region since it can be utilized
to measure the evolving geometry of the interaction zone while
being applied to the studies to search for a possible quark-gluon
plasma and study the properties of the predicted new state of
matter \cite{Heinz,Pratt,Sullivan,Achim}.

In the applications of experimental and theoretical heavy ion
reactions, various aspects have been investigated via the HBT
correlation functions (for the review, e.g., see
Refs.\cite{Boal,Heinz}). Overlooking the literatures
\cite{Tanihata1,Zahar,Arnell,RIA,Li}, more studies on HBT with
unstable nuclear beam can be explored. In terms of the structure
of the nuclei, integral measurements, such as total reaction cross
sections \cite{Shen,Ma1,Ma2,Ozawa}, are only sensitive to the
overall size. However, the dissociation reactions, in which the
core and/or nucleons are detected in the final state, can provide
some structure information \cite{Orr}. The major difficulty is the
relationship between the initial and final states as dictated by
the distorting effects of the reaction \cite{Marques1,Marques2}.
It is very interesting to investigate the  nuclear structure via
HBT technique further.

The binding energy is very important for the structure of nuclei
and it indicates of the stable level of  nucleus and
nucleon-nucleon relationship among  nucleus. It has been  studied
through the density calculation with the Relativistic Mean Field
theory in the past years. In this paper, we will make a
systematical study of the relationship between the binding energy
per nucleon and the strength of neutron-proton correlation
function  at very small relative momentum with help of
Isospin-Dependent Quantum Molecular Dynamics (IDQMD) model which
can describe the reaction dynamics on the event-by-event basis.

\section{MODEL DESCRIPTION}

Firstly, we would like to  recall the HBT technique. As we know
the wave function of relative motion of light identical particle,
when emitted in close proximity in space-time, is modified by the
final-state interaction and quantum statistical symmetries, and
this is the principle of the intensity interferometry, i.e. HBT.
In standard Koonin-Pratt formalism \cite{Koonin,pratt1,pratt2},
the two-particle correlation function is obtained by convoluting
the emission function $g(\mathbf{p},x)$, i.e., the probability for
emitting a particle with momentum $\mathbf{p}$ from the space-time
point $x=(\mathbf{r},t)$, with the relative wave function of the
two particles, i.e.,
\begin{equation}
C(\mathbf{P},\mathbf{q})=\frac{\int d^{4}x_{1}d^{4}x_{2}g(\mathbf{P}%
/2,x_{1})g(\mathbf{P}/2,x_{2})\left| \phi (\mathbf{q},\mathbf{r})\right| ^{2}%
}{\int d^{4}x_{1}g(\mathbf{P}/2,x_{1})\int
d^{4}x_{2}g(\mathbf{P}/2,x_{2})}. \label{CF}
\end{equation}%
where  $\mathbf{P(=\mathbf{p}_{1}+\mathbf{p}_{2})}$ and $\mathbf{q(=}%
\frac{1}{2}(\mathbf{\mathbf{p}_{1}-\mathbf{p}_{2}))}$ are the
total and relative momenta of the particle pair respectively, and
$\phi (\mathbf{q}, \mathbf{r})$ is the relative two-particle wave
function with $\mathbf{r}$
being their relative position, i.e., $\mathbf{r=(r}_{2}\mathbf{-r}_{1}%
\mathbf{)-}$ $\frac{1}{2}(\mathbf{\mathbf{v}_{1}+\mathbf{v}_{2})(}t_{2}-t_{1}%
\mathbf{)}$. This approach has been very useful in studying
effects of nuclear equation of state and nucleon-nucleon cross
sections on the reaction dynamics of intermediate energy heavy-ion
collisions \cite{bauer}. In this work, we use the Koonin-Pratt
method to investigate the neutron-proton correlation function of a
few isotope chains to explore the binding energy dependence of the
HBT strength in intermediate energy heavy-ion collisions.

Using the computation code Correlation After Burner  of Pratt
\cite{crab}, which takes into account final-state nucleon-nucleon
interactions, we have constructed two-nucleon correlation
functions from the emission phase space  given by the IDQMD model.

Interpretation of correlation functions measured in heavy-ion
collisions requires understanding the relationship of the
parameters extracted from fitting the data and the true
single-particle distributions at the freeze-out. This relationship
can be established by using an event generator that models the
collision dynamics, particle production, and then construct a
two-particle correlation function. The event-generator correlation
functions are constructed from the positions and momenta
representing the single-particle emission distribution at the time
of the last strong interaction, i.e. at freeze-out. The
event-generator used in our work is the Isospin-dependent Quantum
Molecular Dynamics transport model, which has been applied
successfully to the studies of the heavy-ion collisions process
\cite{Aichelin,Ma3,Zhang}. For completeness, we would like to give
a brief sketch of the model in following.

The Quantum Molecular Dynamics (QMD) approach  is an n-body theory
to describe heavy ion reactions from intermediate energy to 2
GeV/n. It includes five important parts: the initialization of the
target and the projectile; the propagation in the effective
potential; the collisions between the nucleons; the Pauli blocking
effect and the numerical tests. A general review about the QMD
model can be found in \cite{Aichelin}.

As we know, the dynamics in heavy-ion collisions at intermediate
energies is mainly governed by three components, namely the mean
field, two-body collisions, and Pauli blocking. Therefore, for an
isospin-dependent reaction dynamics model it is important for
these three components to include isospin degrees of freedom. What
is more essential, in the initialization of projectile and target
nuclei, the samples of neutrons and protons in phase space should
be treated separately because there exists a large difference
between neutron and proton density distributions for nuclei far
from the $\beta$-stability line. Particularly, for light
neutron-rich nucleus one should sample a stable initialized
nucleus with neutron-skin structure and therefore one can directly
explore the nuclear structure effects through a microscopic
transport model. The IDQMD model has been improved based on the
above ideas.

\begin{figure}
\begin{center}
\includegraphics[scale=0.5]{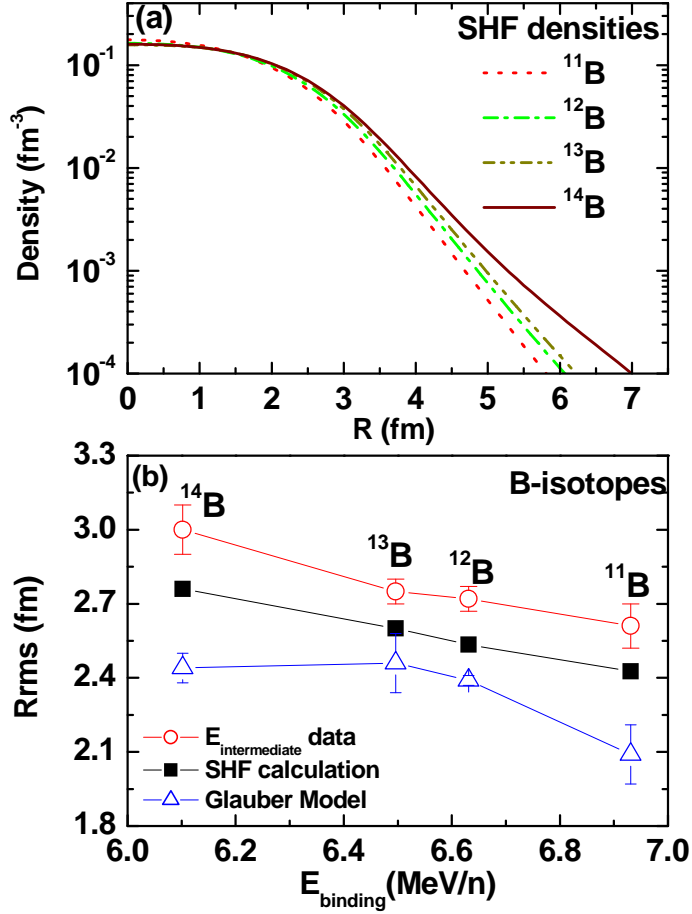}
\end{center}
\caption{\label{Fig_dens}(a) The total density distributions of
$^{11-14}B$ nuclei calculated from the SHF model with parameter
set SKM*. (b) The RMS radii of $^{11-14}B$ nuclei as a function of
the binding energy. The solid squares are the SHF results from
Fig.~\ref{Fig_dens}(a), the open circles are the data from
intermediate energy reactions \cite{Tani} and the open triangles
represent the data with the Glauber model calculation with a
optical limit approximation \cite{Opt}.}
\end{figure}

The nuclear size and density distribution are important bulk
properties of nuclei that are mainly determined by the nuclear
potential, single-particle orbits and wave function \cite{Ozawa1}.
In the IDQMD model, the neutrons and protons are distinguished
from each other in the initialization of projectile and target
nuclei. The references of neutron and proton density distributions
for the initial projectile and target nuclei in IDQMD are taken
from the Skyrme-Hartree-Fock (SHF) method with parameter set SKM*
and they can  be obtained through Monte-Carlo sampling technique
in IDQMD which gives the similar distributions to the input ones
from SHF. The input density distributions of B-isotopes which are
used as the references of the IDQMD initialization of the
projectiles by Monte-Carlo sampling in our calculation are shown
in Fig.~\ref{Fig_dens}(a). From the figure, we can see, with the
increasing mass number, the total density distribution extends a
larger range. In general, it is known that the radius of the
isotopes becomes larger with the rising of neutron number. Our
calculated density distribution reproduces this tendency very
well. The relationship between the RMS radius and the binding
energy per nucleon is shown in Fig.~\ref{Fig_dens}(b). It is
obvious that the radius tends to be small with the increasing
binding energy per nucleon. In this figure, we plot three-set
values of RMS radius, namely SHF radius from our calculations
(solid squares), the experimental data \cite{Tani} and the deduced
values from Glauber model \cite{Opt}. As we know, the binding
energy per nucleon indicates of the compactness among nucleons.
While, HBT can give us the spatial information, such as source
size. Therefore it is very interesting to investigate the
tightness of the nuclei through HBT. With these suitable nuclear
density distributions we could get the coordinate of nucleons in
the initial nuclei in terms of the Monte-Carlo sampling method.
The momentum distribution of nucleons is generated by means of the
local Fermi gas approximation.

\begin{figure}
\vspace{-0.6truein}
\begin{center}
\includegraphics[scale=0.4]{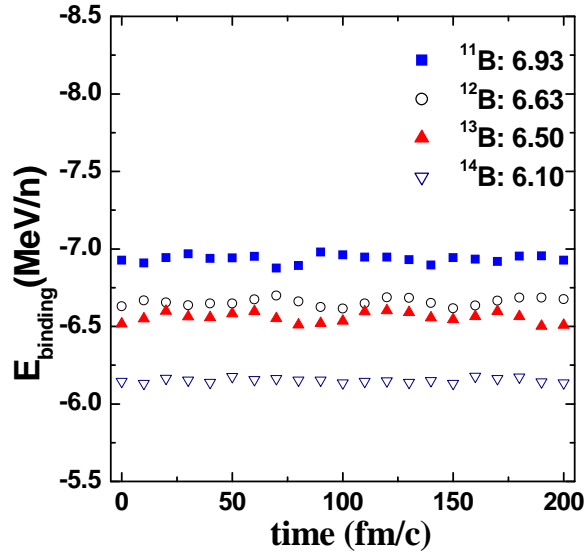}
\end{center}
\vspace{-0.7truein} \caption{ \label{Fig_Ebind}The time evolution
of the binding energy per nucleon for $^{11-14}B$ isotopes. The
noted numbers in the figure are the experimental value of the
binding energy per nucleon. }
\end{figure}

It is well known that the binding energy ( $E_{binding}$ ) of the
nuclei incarnates the stability level of the nucleus  and the
relationship among the nucleons. With higher $E_{binding}$ per
nucleon, normally, the nuclear stability is higher too. It is an
important way to study the inner structure of the nuclei. In some
simulation models $E_{binding}$ is considered as a benchmark so
that the most suitable results can be obtained. In this paper we
shall examine its systematics through HBT approach with the
event-generator IDQMD model.

In the IDQMD model in zero temperature, $E_{binding}$ per nucleon
of the nuclei can be obtained and its stability can be checked
till a longer time. During the evolution process lasted till 200
fm/c, the transport process of the nucleons are governed by the
initialization,  the Pauli blocking effects, nucleon-nucleon
collisions and nuclear mean field etc. In our calculations, only
those events which keep good stability according to their radii
and $E_{binding}$ per nucleon are selected. From these events we
can choose the most suitable initialization phase space. For
instance, for isotopes $^{11-14}B$ which we will use in our
simulation calculation, the time evolution of the binding energy
in zero temperature  are shown in Fig.~\ref{Fig_Ebind}. The dots
are the simulation results of the binding energy per nucleon of
the isotopes with the time. The numbers behind the labels are the
experimental $E_{binding}$ values of the isotopes. From
Fig.~\ref{Fig_Ebind}, it is clearly that the  values of the
binding energy per nucleon are almost the same as  the
experimental value and the nucleus  keeps its own stability
throughout the simulation process in the case of no thermal
excitation.

\section{RESULTS AND DISCUSSION}

The approach of HBT with IDQMD has been applied in reproducing the
halo neutron-halo neutron correlation function of HBT results in
IDQMD \cite{Wei}. In our work \cite{Wei}, the nucleons are defined
as emitted if they do not belong to any clusters ($A \geq$ 2)
which are recognized by a simple coalescence model: i.e. nucleons
are considered  to be part of a cluster if in the end at least one
other nucleon is closer than $r_{min} \leq $ 3.5 fm in coordinate
space and $p_{min} \leq$ 300MeV/c in momentum space in the final
state. With this approach, the experimental data could be
reproduced well.

Based on the above success to fit the data, we further made a
prediction for the $E_{binding}$ dependence of n-p HBT strength
for $Be$-isotopes in \cite{Wei}. Here we wonder its applicability
for other light isotopes. To the end,  we select that the target
is $^{12}C$ and the projectiles are $Li$,  $B$, $N$ and
$C$-isotopes, respectively. The incident energy is 800 MeV/n. The
emitted nucleons which are taken into account in the HBT
calculation are all from the projectiles.

\begin{figure}
\begin{center}
\vspace{-0.1truein}
\includegraphics[scale=0.5]{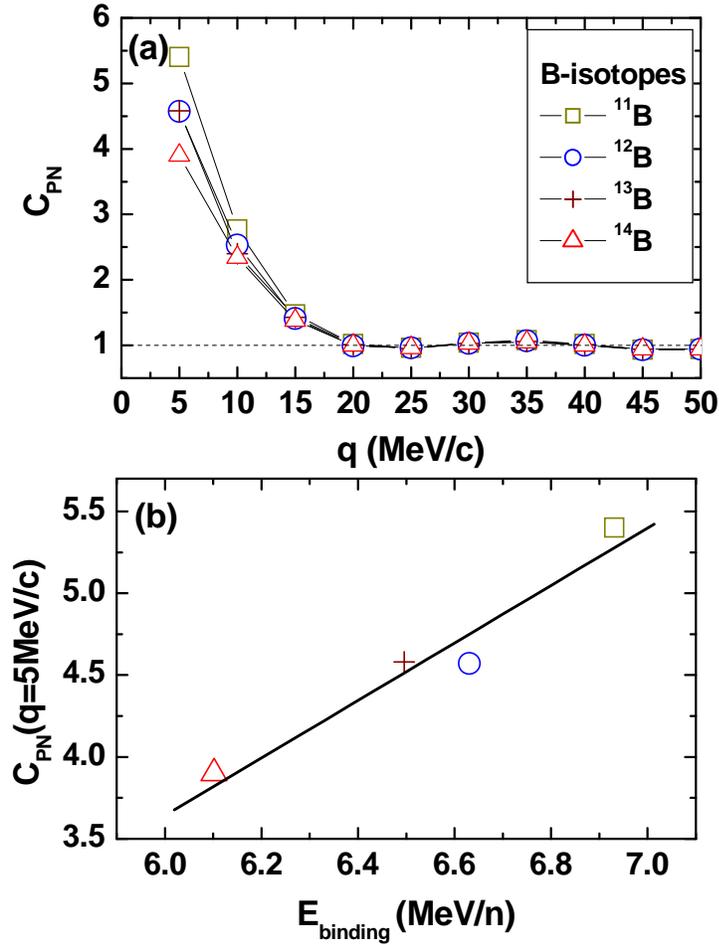}
\end{center}
\vspace{-0.1truein} \caption{\label{Fig_HBT_B}  The proton-neutron
correlation function for different $B$ isotopes (a) and the
relationship between the strength of proton-neutron correlation
function $C_{PN}$ at 5 MeV/c and the binding energy per nucleon of
the projectile $E_{binding}$ (b). The collisions were simulated at
800 MeV/n of the incident energy and head-on collisions. The
target is $^{12}$C.}
\end{figure}

The calculated results are shown in Fig.~\ref{Fig_HBT_B} at 200
fm/c which is well later than the freeze-out time for the central
collisions at  800 MeV/n. The soft potential (an incompressibility
of $K$ = 200 MeV) and the asymmetrical strength $C_{sym}$  = 32
MeV between the neutron and the proton is used in IDQMD
simulation. Fig.~\ref{Fig_HBT_B}(a) shows the n-p correlation
function which peaks at $q \approx 0$ \textrm{MeV/c}. The
deviation of $C(q)$ from 1 thus reflects the information of the
emission source. Fig.~\ref{Fig_HBT_B}(b) shows the relationship
between the strength of n-p correlation function $C_{PN}$ at 5
MeV/c  versus the binding energy per nucleon of the projectile
$E_{binding}$. The solid line is just a linear fit to guide the
eyes.

From the figure, it is clearly that the strength of HBT at very
small relative momentum shows a distinct dependence on the neutron
number and the tendency of $C_{PN}$ rises with the increasing of
$E_{binding}$. Among the projectiles we studied, the number of the
protons is five and that of the neutron is gradually increasing.
With the increasing of the mass number, the mean relationship
among the nucleons changes weaker in general. Since the strength
$C_{PN}$ of correlation function reflects the mean relationship
among the emitted neutrons and protons and the binding energy
associates with the tightness among the nucleons, it is reasonable
to exist such a relationship between these two parameters.
Actually, the tendency shown in Fig.~\ref{Fig_HBT_B} indicates
that $C_{PN}$ can reveal the compactness of the nuclei.

\begin{figure}
\begin{center}
\vspace{-0.9truein}
\includegraphics[scale=0.5]{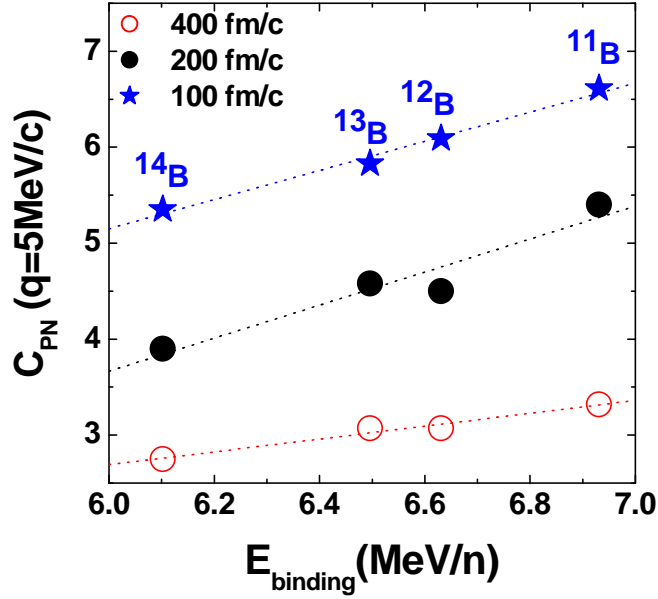}
\end{center} \vspace{-0.9truein} \caption{\label{Fig4} The
comparison of the n-p HBT strength at different evolution time.
The stars, filled circles and opened circles are the results at
the evolution time of 100fm/c, 200fm/c and 400fm/c. The dotted
lines are linear fits to guide the eyes. }
\end{figure}

Considering that the correlation function is influenced by the
spatio-temporal information of the emitted particles, it is
necessary to investigate the HBT results in different evolution
time. The results are shown in Fig.~\ref{Fig4}. For our systems at
800 MeV/n the freeze-out time is less than 100 fm/c in IDQMD
model. From Fig.~\ref{Fig4} one can find that the HBT strength has
the strong dependence with the $E_{binding}$ at the different
evolution time. Generally speaking, the strength of HBT becomes
weak in late stage of time evolution which is a natural result
since the system tends more dilute which reduces the spatial
correlation between nucleons. However, the tendency of
relationship between the HBT strength and the $E_{binding}$ does
not change despite the different value of $C_{PN}$ reveals when
the different evolution time is used. As a result, it is reliable
to investigate qualitatively the behavior of HBT strength at a
certain evolution time which is later than the freeze-out one.
This is a reason why we choose t = 200 fm/c to extract the
correlation function as shown in Fig.~\ref{Fig_HBT_B}.

\begin{figure}
\begin{center}
\vspace{-0.1truein}
\includegraphics[scale=0.5]{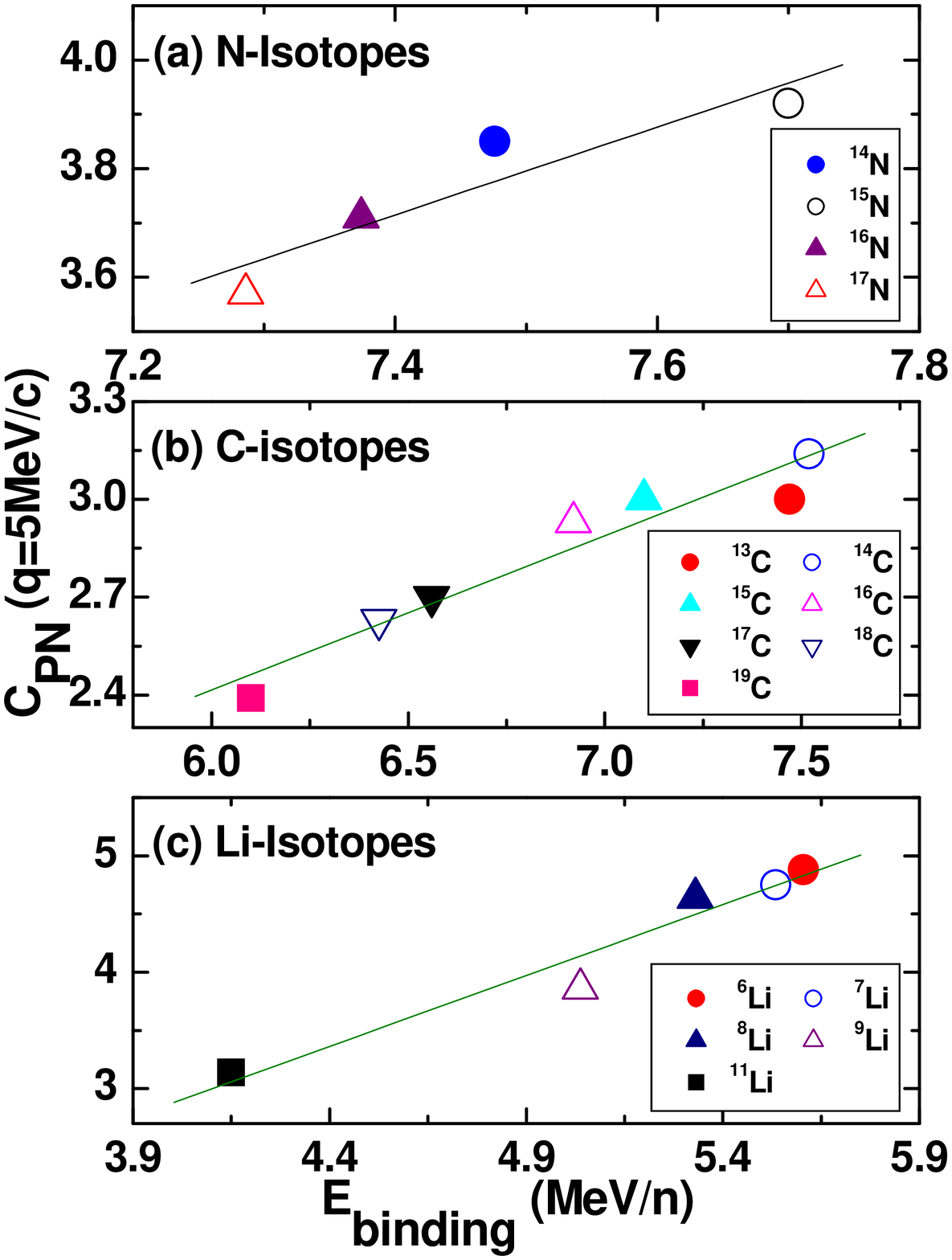}
\end{center}
\vspace{0.1truein} \caption{\label{Fig5} The similar results with
Fig.3(b) but for different isotopes.}
\end{figure}

In addition, we investigated the relationship of $C_{PN}$ versus
$E_{binding}$ in other three isotopes of $Li$, $C$ and $N$. The
calculated results are shown in Fig.~\ref{Fig5}. Again the HBT
results are extracted at t = 200 fm/c. The calculation condition
is the same as that of $B$-isotopes. From Fig.~\ref{Fig5} one can
find that the nearly linear relationship between the $E_{binding}$
per nucleon and the strength of proton-neutron correlation
function $C_{PN}$ at 5 MeV/c exists in these different isotopes.

\section{SUMMARY}

In summary, the intensity interferometry technique has been used
to investigate systematically  its sensitivity to the binding
energy of light nuclei from the break-up of nuclei by convoluting
the phase-space distribution generated with the IDQMD model.
Firstly, we use SHF to determine the reasonable initial density
distributions which are taken as the references of IDQMD
initialization of projectile and target by the Monte-carlo
sampling. The stability has been checked by the  time evolution of
binding energy in IDQMD at zero temperature and the good stable
initial phase space has been selected for the simulation of the
event-generator IDQMD model afterwards. The dependence of the
proton-neutron correlation functions ($C_{PN}$) at small relative
momentum with the binding energy ($E_{binding}$) for four isotope
chains ($Li$, $B$, $C$ and $N$) has been systematically explored.
It was found that the correlation strength of $C_{PN}$ at small
relative momentum rises with the binding energy per nucleon. It is
of particular interest if one can explore experimentally.

\ack
This work was supported in part by the Major State Basic
Research Development Program under Contract No G200077404, the
Chinese Academy of Sciences Grant for the Distinguished Young
Scholars of National Natural Science Foundation of China (NNSFC)
under Grant No. 19725521, NNSFC under Grant Nos 10135030 and
10328509.

{}


\section*{References}
\begin{thebibliography}{}

\bibitem{Brown} Hanbury Brown R, Twiss R Q 1956 {\it Nature} {\bf 178}
1046.
\bibitem{Goldhaber}  Goldhaber G {\it et al} 1960 {\it Phys. Rev.} {\bf 120} 300
\bibitem{Koonin} Koonin S E  1977 {\it Phys. Lett.} B {\bf 70}  43
\bibitem{Boal} Boal D H{\it et al} 1990 {\it  Rev. Mod. Phys.} {\bf 62} 553
\bibitem{Heinz} Heinz U{\it et al} 1999 {\it Annu. Rev. Nucl. Part. Sci.} {\bf 49} 529
\bibitem{Pratt} Pratt S 1984  {\it Phys. Rev. Lett.} {\bf 53} 1219
\bibitem{Sullivan} Sullivan J P {\it et al} 1993 {\it Phys. Rev. Lett} {\bf 70} 3000
\bibitem{Achim} Achim U, Heinz U 1997 {\it Phys. Rev.} C {\bf 56} 610
\bibitem{Tanihata1}  Tanihata I {\it et al} 1985 {\it Phys. Lett.} B {\bf 160} 380
\bibitem{Zahar}  Zahar M {\it et al} 1993 {\it Phys. Rev.} C {\bf 48} R1484
\bibitem{Arnell} Anne R {\it et al} 1990 {\it   Phys. Lett.} B {\bf 250} 19
\bibitem{RIA} RIA Physics White Paper 2000 edited by R. Casten {\it et
al}
\bibitem{Li} Li B A {\it et al} 2004 {\it  Nucl. Phys.} A {\bf 735}   563
\bibitem{Shen} Shen W Q {\it et al} 1989 {\it Nucl. Phys.} A {\bf 491}  130
\bibitem{Ma1} Ma Y G {\it  et al} 1993 {\it Phys. Lett.} B {\bf 302} 386
           \\ Ma Y G {\it  et al} 1993 {\it Phys. Rev.} C {\bf 48}  850
\bibitem{Ma2} Ma Y G, Zhang H Y and Shen W Q 2002 {\it Prog. Phys. (in Chinese)} {\bf 22} 99
\bibitem{Ozawa} Ozawa A {\it et al} 1996 {\it Nucl. Phys.} A {\bf 608} 63
\bibitem{Orr} Orr N A 1997 {\it Nucl. Phys.} A {\bf 616}  155
\bibitem{Marques1} Marques F M {\it et al} 2000 {\it Phys. Lett.} B {\bf 476} 219
\bibitem{Marques2} Marques F M {\it et al} 2001 {\it Phys. Rev.} C {\bf 64} 061301

\bibitem{pratt1} Pratt S 1986  {\it Phys. Rev.} D \textbf{33} 72

\bibitem{pratt2} Pratt S, Tsang  M B 1987 {\it Phys. Rev.} C {\bf 36}
2390

\bibitem{bauer} Bauer W,  Gelbke C K,  Pratt S 1992 {\it  Ann. Rev. Nucl.
Part. Sci.} {\bf 42} 77

\bibitem{crab} Pratt S 1994 {\it  Nucl. Phys.} {\bf A 566}  103c

\bibitem{Aichelin} Aichelin J 1991 {\it Phys. Rep.} {\bf 202} 233
\bibitem{Ma3} Ma Y G, Shen W Q 1995 {\it Phys. Rev.} C {\bf 51} 710
\bibitem{Zhang} Zhang F S {\it et al} 1999 {\it Phys. Rev.} C {\bf 60}  064604

\bibitem{Ozawa1} Ozawa A, Suzuki T, Tahihata I 2001 {\it Nucl. Phys.} A {\bf 693} 32

\bibitem{Tani}Tanihata I {\it et al} 1988 {\it Phys. Lett.} B {\bf 206}  592


\bibitem{Opt} Liatard E {\it et al} 1990 {\it Eurphys. Lett.} {\bf 13} 401
\bibitem{Wei} Wei Y B, Ma Y G, Shen W Q, Ma G L, Wang K, Cai X Z, Zhong C,
Guo W, Chen J G 2004  {\it Phys. Lett.} B {\bf 586} 225.

\end{thebibliography}
\end{document}